\documentclass[aps,prl,
reprint,
letterpaper,
groupedaddress,
twocolumn,
floatfix,
amsmath,amssymb,amsfonts,
]{revtex4-1}

\usepackage{graphicx}
\usepackage[breaklinks=true]{hyperref}
\usepackage{color}
\usepackage{array}
\usepackage[normalem]{ulem}

\begin{document}

\title{Accelerating the convergence of replica exchange simulations\\ using Gibbs sampling and adaptive temperature sets}

\author{Thomas Vogel} 
\email{tvogel@lanl.gov; tvogel1@stetson.edu}
\author{Danny Perez}
\email{danny\_perez@lanl.gov}

\affiliation{Theoretical Division (T-1), Los Alamos National
  Laboratory, Los Alamos, NM 87545}

\begin{abstract}
  We recently introduced a novel replica-exchange scheme in which an
  individual replica can sample from states encountered by other
  replicas at any previous time by way of a global configuration
  database, enabling the fast propagation of relevant states through
  the whole ensemble of replicas. This mechanism depends on the
  knowledge of global thermodynamic functions which are
  \textit{measured} during the simulation and not coupled to the heat
  bath temperatures driving the individual simulations.  Therefore,
  this setup also allows for a continuous adaptation of the
  temperature set. In this paper, we will review the new scheme and
  demonstrate its capability. The method is particularly useful for
  the fast and reliable estimation of the microcanonical temperature
  $T(U)$ or, equivalently, of the density of states $g(U)$ over a wide
  range of energies.
 
\end{abstract}

\maketitle

\section{Introduction}
\label{intro}

%\enlargethispage{\baselineskip}
Simulating physical systems exhibiting multifaceted structural
transitions (encountered, for example, in many protein folding
problems) or containing extremely hard to find but thermodynamically
important states (e.g., crystalline states in hard materials) is a
notorious challenge.  Replica-exchange (RE) and multicanonical (MUCA)
sampling schemes have both been applied, in conjunction with Monte
Carlo (MC) and molecular dynamics (MD) methods, to address this issue.
They enable the investigation of systems at multiple or varying
temperatures, such facilitating the exploration of the configurational
space. One of the main practical challenges in the case of RE is the
determination of good temperature sets for the heat baths. To this
day, many sophisticated schemes have been proposed to solve this
problem, see,
e.g.,~\cite{rathore05jcp,katzgraber06jsm,patriksson08pccp,guidetti2012jcp,ballard14jctc}.
Another limitation of generic RE schemes is that exchange of
configurations obviously does not, on its own, change the ensemble of
configurations present in the simulation. This is not optimal if the
simulation is not converged yet, e.g., if some replicas have yet to
locate all thermodynamically relevant states. In this case, some sort
of population control, where ``good'' configurations can be multiplied
and ``irrelevant'' ones deleted, can help
(see~\cite{hsugrassberger11jsp} for an example).

MUCA simulations offer a different strategy to efficiently sample
configuration space. In the case of MD, MUCA sampling simply consists
of rescaling the interatomic forces:
$f_\mathrm{muca}=(T_0/T(U))\,f_\mathrm{can}$~\cite{hansmann96cpl,junghans14jctc},
where $f_\mathrm{can}=-\nabla U$ are the conventional forces and
$T(U)$ is the microcanonical temperature.  The difficulty lies in the
estimation of this \textit{a priori} unknown function.
Metadynamics~\cite{laioparrinello02pnas,dama14prl} and statistical
temperature molecular dynamics (STMD)~\cite{kim06prl} were
successfully applied to this problem.  However, these methods can show
dynamical artifacts like hysteresis when the system is driven over
first-order like transitions, for example. Further, even with the
knowledge of $T(U)$, the diffusion of a MUCA simulation in energy
space can be very slow. However, a very powerful side-effect of MUCA
simulations is that, once obtained, $T(U)$ can be used to infer the
density of states, hence providing a wealth of thermodynamic
information.

In the following, we review a generalized RE method, the
Gibbs-sampling enhanced RE, which addresses some of the issues with
generic RE and MUCA simulations. By uncoupling the measured
microcanonical temperature from the heat-bath temperatures which drive
the simulations we can introduce powerful conformational mixing steps
and continuously adapt the heat-bath temperatures for a complete
coverage of the whole energy range. We demonstrate this ability by
presenting a particular adaptive simulation which is set up so that
only a tiny part of the phase space in covered the beginning.  Note
that $T(U)$ is also available at the end of such a simulation,
providing an alternative to conventional MUCA approaches.

\section{Method review}

We consider a RE scheme~\cite{geyer91proc} in an expanded-canonical
ensemble~\cite{lyubartsev92jcp} with reference temperatures $T_0^i$.
In contrast to other sampling methods where thermodynamic functions
are continuously
adapted~\cite{wanglandau01prl,laioparrinello02pnas,junghans14jctc},
the basic idea in our method~\cite{vogelperez2015prl} is to
\textit{measure} the microcanonical temperature while running
canonical simulations, hence decoupling the measurements from the
simulation-driving heat bath temperatures. We use an estimator for the
microcanonical temperature based the microcanonical averages of
time-derivatives of the product between the normalized energy gradient
$\eta=\nabla U/(\nabla U\nabla U)$ and the particle momenta $p$:
\begin{equation}
\label{eq:g}
g=\frac{\mathrm{d} F(U)}{\mathrm{d}
  U}=-\left\langle\frac{\mathrm{d}}{\mathrm{d}t}(\eta \cdot p)\right\rangle\,,
\end{equation}
where $F$ is the free energy, a relation which was derived in a more
general form in the context of the adaptive biasing force (ABF)
method~\cite{darve08jcp}. The time derivatives are estimated in an
additional, constant-energy MD time step. As defined,
$g$ relates to the microcanonical temperature as:
\begin{equation}
\label{eq:TABF}
T_\mathrm{ABF}(U)=\frac{T_0}{1-g}\,.
\end{equation}
% Alternative estimators also exist, see~\cite{rugh97prl,butler98jcp}
% for example.
During the simulation, all replicas contribute measurements to the
global calculation of $T_\mathrm{ABF}(U)$, while they also
collectively fill a global conformational database.\footnote{The
  management of all global data is done by an additional master
  process which does not perform a MD run himself.}  An instantaneous
estimator for the (unnormalized) entropy $S(U)$ can then be obtained:
\begin{equation}
\label{eq:SE}
S(U)=\int_{U_0}^U T^{-1}_\mathrm{ABF}(U^\prime)\,\mathrm{d}U^\prime+C\,.
\end{equation}
With $g(U)=\mathrm{exp}[k_B^{-1}S(U)]$ being the density of states,
canonical energy probability distributions $P_i(U)\propto
g(U)\,\mathrm{e}^{-\beta_i U}$ can be calculated.  The key is that
these distributions calculated from global data are different from
those that would be inferred from the data gathered by individual
walkers alone. For example, $P_i(U)$ can account for the presence of
some low energy state, even though walker $i$ might never have sampled
it himself. This allows for the introduction of a rejection-free,
global Monte Carlo move facilitating the propagation of important
states through the simulation: at any time, a walker can sample a new
energy value from the distribution $P_i(U)$ (a so-called ``Gibbs
sampling'' move) and replace its actual configuration by one with that
energy ($\pm\epsilon$, for practical reasons) from the global
configuration database. Such moves allow each replica to properly
sample from the pool of states visited by all other replicas.

\section{The adaptive temperature set}

In RE, good performance relies on a proper choice of temperatures,
as this determines the {exchange} probabilities between neighboring
replicas, say 
$i$ and \hbox{$i+1$}. Assuming the corresponding canonical distributions
$P_i(U)$ and $P_{i+1}(U)$ are known, the {exchange} probabilities
$W_i(P_i,P_{i+1})$ read:
\begin{align}
  \label{eq:Ws}\nonumber
  &W_i(P_i,P_{i+1})\\
  &=\int_{-\infty}^\infty P_i(U)
  \int_{-\infty}^\infty W_{\mathrm{acc}}(U,U^\prime)
  P_{i+1}(U^\prime)\, \mathrm{d}U \mathrm{d}U^\prime\,,
\end{align}
with
\begin{equation}
  \label{eq:Wacc}
  W_{\mathrm{acc}}(U,U^\prime)=\min\left\{1,\mathrm{exp}\left[(\beta_{i+1}-\beta_i)(U^\prime-U)\right]\right\}
\end{equation}
being the RE acceptance probability.  These $W_i(P_i,P_{i+1})$ can be
approximated (see~\cite{nadler07pre}, for example), but with the
knowledge of a global, instantaneous $T(U)$ and hence of $g(U)$ and
$P_i(U)$, the $W_i(P_i,P_{i+1})$ can be \textit{calculated}. We can
then find a temperature set $\beta^{-1}_i=k_BT_0^i$, $(i=1,\ldots,n)$,
such that [cf.~\cite{nadler07pre,elmar08prl}]
\begin{equation}
  \label{eq:Wconst}
  W_i(P_i,P_{i+1})=\mathrm{const}\quad\forall i\,.
\end{equation}
Such a criterion would be optimal in the limit where sampling is
ergodic on the timescale of exchange
attempts~\cite{nadler07pre,elmar08prl}. When this condition is
broken, it might instead be preferable to directly minimize the
tunneling time ~\cite{katzgraber06jsm}.

Since there is no need for the reference temperatures $T_0^i$ to
remain constant for the calculation of $T_\mathrm{ABF}(U)$ or other
thermodynamic averages, these can be freely changed during the
simulation.  With a given, fixed number of processors, there are
two options: either require a fixed value for the constant in
Eq.~\ref{eq:Wconst} (resulting in a fluctuating total temperature
range) or fix the global temperature range (and thus correspondingly
adjusts the value of the constant). We chose the latter to ensure we
always cover the whole temperature range and adjust the temperatures
using a bisection scheme.

\section{Demonstration}

\begin{figure}[b!]
\includegraphics[width=\columnwidth]{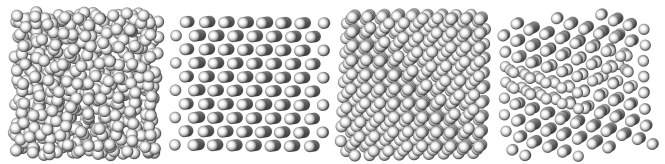}
\caption{Representative configurations of the 500-silver atom system. From
  left to right: an amorphous state; the pure crystalline (ground-)
  state; two crystalline states with stacking faults.
 \label{fig:configs}}
\end{figure}
\begin{figure*}
\includegraphics[width=\textwidth]{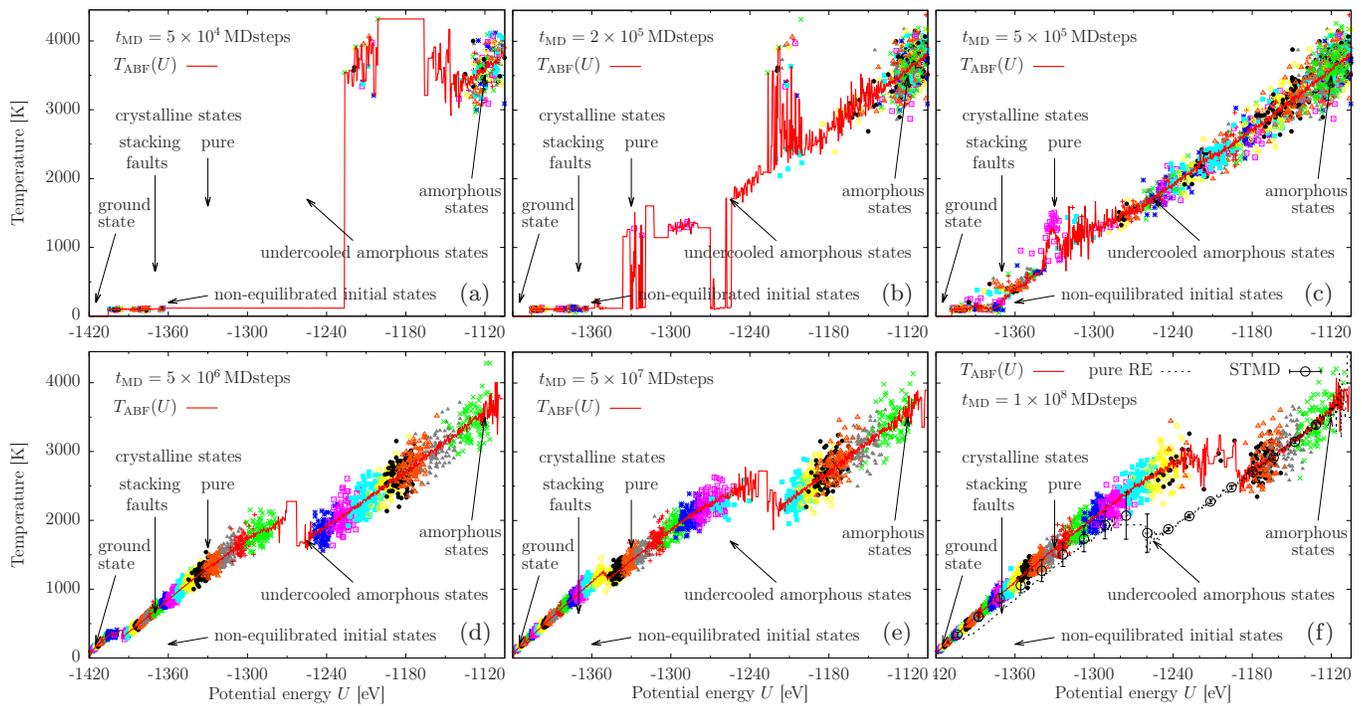}
\caption{Snapshots of the instantaneous estimator for the
  microcanonical temperature (red, solid line) at different times
  during the simulation. The dots correspond to raw measurements
  entering in the average (cf. Eqs.~\ref{eq:g} and~\ref{eq:TABF}).
  Different colors and symbols correspond to different reference
  temperatures $T_0^i$. Raw measurements are only shown for every
  other reference temperature. Only 10\% of the raw data are shown for
  clarity. {The last plot (f) also contains data from pure RE and
  from STMD (see text for details)}.\label{fig:snapshots}}
\end{figure*}
\begin{figure*}
\centering
\includegraphics[width=.86\textwidth]{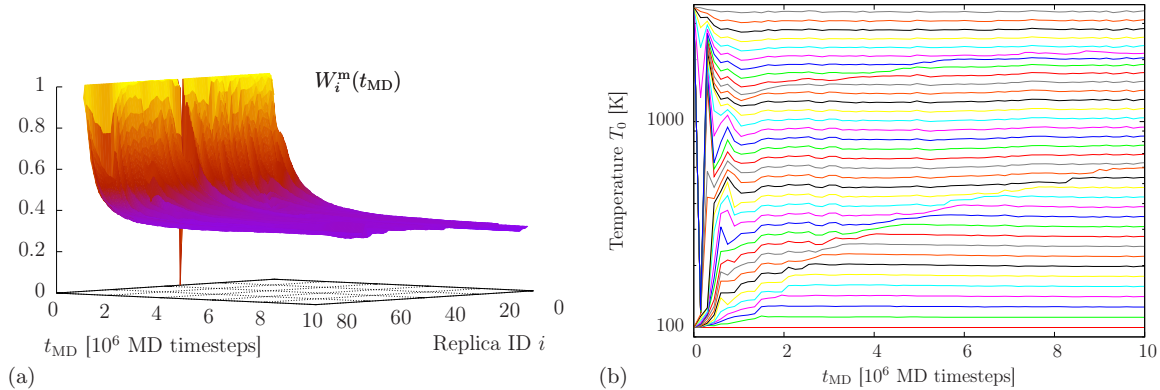}
\caption{(a) Measured replica-exchange acceptance rates for all
  neighbor pairs ($i,i+1$) over time. (b) Time evolution of the temperature set
  $T_0^i$.\label{fig:WT}}
\end{figure*}
We illustrate the performance of the method for a system of 500 silver
atoms interacting via an embedded atom
potential~\cite{williams06msmse}. We use a constant particle density
of $\rho=0.0585\AA^{-3}$ and perform a Gibbs-sampling enhanced RE
simulation based on molecular dynamics runs at heat-bath temperatures
$T_0^i$.  Representative configurations are shown in
Fig.~\ref{fig:configs}.  We begin the simulation in an extremely
unnatural way: of the 71 individual canonical walkers, half of them
start at the minimum temperature of $T_0^{i<35}=T_\mathrm{min}=100$\,K
and the other half at the maximum temperature
$T_0^{i\geq35}=T_\mathrm{max}=3500$\,K.  Furthermore, all replicas
begin from an amorphous configuration.  Figure~\ref{fig:snapshots}
shows snapshots of the instantaneous microcanonical temperature
measured as $T_\mathrm{ABF}(U)$ (cf.  Eq.~\ref{eq:TABF}) for different
simulation times $t_\mathrm{MD}$. Shortly after starting the
simulation (Fig.~\ref{fig:snapshots}\,a) and before the first
temperature adaptation, we only find configurations with temperatures
scattered around the initial values $T_\mathrm{min}$ and
$T_\mathrm{max}$. The instantaneous temperature $T_\mathrm{ABF}(U)$ is
completely arbitrary at this point.\footnote{If there are no measured
  values in particular energy bins, we simply copy the closest valid
  value from a lower energy.}  The next snapshot (b) was taken after
the first temperature adaptation and the exploration of configurations
at intermediate energies starts.\vadjust{\break} After about $5\times10^5$
MD steps (Fig.~\ref{fig:snapshots}\,c; temperature adaptation takes
place every $1.5\times10^5$ MD steps) temperatures are already spread
out such that the whole energy range is covered.  The ground state has
been discovered after $5\times10^6$ MD steps (d) and propagates
through the simulation (e) facilitated by the Gibbs sampling moves.
The simulation eventually converges after approximately $1\times10^8$
MD steps (f). Note that we set a memory time after which data is
discarded, so that non-equilibrated states encountered in the
beginning do not enter in the averages (Eqs.~\ref{eq:g}
and~\ref{eq:TABF}) at later times.

We show in Fig.~\ref{fig:WT} the \textit{measured} replica-exchange
acceptance rate $W_i^\mathrm{m}$ (a) and the values of the heat bath
temperatures $T_0^i$ (b) as the simulation time progresses. In
accordance to the discussion above, $W_i^\mathrm{m}=0$ for the
exchange between the two neighboring walkers at $T_\mathrm{min}$ and
$T_\mathrm{max}$ and $W_i^\mathrm{m}=1$ everywhere else. After a short
time, however, all measured acceptance rates approach a constant value
of $W_i^\mathrm{m}\approx0.35$. This value is in agreement with the
calculated value for the constraint given in Eq.~\ref{eq:Wconst} for
our particular set-up, i.e., for the fixed temperature range and the
number of walkers we employ. Note that we can directly control the
value of $W_i^\mathrm{m}$ by deploying more or less walkers.
The adaptation speed of the replica-exchange rates is of course
related to the adaption speed of the temperature set $T_0^i$. It can be
seen in Fig.~\ref{fig:WT}\,b that very few temperature adaptation steps
are required before they are reasonably well distributed. Adaptation is
marginal at times $t_\mathrm{MD}\gtrsim1\times10^6$ and is mainly in
response to the propagation of the ground-state, as this creates an
artificial signal of a ``phase transition'' in $T_\mathrm{ABF}(U)$ and
the temperatures get locally closer together around the corresponding
``transition temperature''.

Finally, in Fig.~\ref{fig:snapshots}\,f, we also provide data from a
pure RE exchange run (i.e., w/o Gibbs moves) at the same time and from
multiple Gauss-kernel STMD~\cite{junghans14jctc} runs, for comparison.
While the pure RE run did not converge at all after this time and
mainly samples configurations with stacking faults in the crystalline
phase, the STMD run samples both, pure and defected states but suffers
from hysteresis\footnote{This behavior is known and simply illustrates
  the perils of choosing a reaction variable that involves
  first-order-like transitions.}.  While both conventional methods are
not able to converge in this challenging situation, our method
including additional Gibbs-sampling moves works remarkably well.

\begin{acknowledgments}
We acknowledge funding by Los Alamos National Laboratory's
(LANL) Laboratory Directed Research and Development ER program. LANL
is operated by Los Alamos National Security, LLC, for the National
Nuclear Security Administration of the U.S. DOE under Contract
DE-AC52-06NA25396. Assigned LA-UR-15-21704.
\end{acknowledgments}

\bibliography{CSP2015_VOGEL}

\begin{thebibliography}{22}%
\makeatletter
\providecommand \@ifxundefined [1]{%
 \@ifx{#1\undefined}
}%
\providecommand \@ifnum [1]{%
 \ifnum #1\expandafter \@firstoftwo
 \else \expandafter \@secondoftwo
 \fi
}%
\providecommand \@ifx [1]{%
 \ifx #1\expandafter \@firstoftwo
 \else \expandafter \@secondoftwo
 \fi
}%
\providecommand \natexlab [1]{#1}%
\providecommand \enquote  [1]{``#1''}%
\providecommand \bibnamefont  [1]{#1}%
\providecommand \bibfnamefont [1]{#1}%
\providecommand \citenamefont [1]{#1}%
\providecommand \href@noop [0]{\@secondoftwo}%
\providecommand \href [0]{\begingroup \@sanitize@url \@href}%
\providecommand \@href[1]{\@@startlink{#1}\@@href}%
\providecommand \@@href[1]{\endgroup#1\@@endlink}%
\providecommand \@sanitize@url [0]{\catcode `\\12\catcode `\$12\catcode
  `\&12\catcode `\#12\catcode `\^12\catcode `\_12\catcode `\%12\relax}%
\providecommand \@@startlink[1]{}%
\providecommand \@@endlink[0]{}%
\providecommand \url  [0]{\begingroup\@sanitize@url \@url }%
\providecommand \@url [1]{\endgroup\@href {#1}{\urlprefix }}%
\providecommand \urlprefix  [0]{URL }%
\providecommand \Eprint [0]{\href }%
\providecommand \doibase [0]{http://dx.doi.org/}%
\providecommand \selectlanguage [0]{\@gobble}%
\providecommand \bibinfo  [0]{\@secondoftwo}%
\providecommand \bibfield  [0]{\@secondoftwo}%
\providecommand \translation [1]{[#1]}%
\providecommand \BibitemOpen [0]{}%
\providecommand \bibitemStop [0]{}%
\providecommand \bibitemNoStop [0]{.\EOS\space}%
\providecommand \EOS [0]{\spacefactor3000\relax}%
\providecommand \BibitemShut  [1]{\csname bibitem#1\endcsname}%
\let\auto@bib@innerbib\@empty
%</preamble>
\bibitem [{\citenamefont {Rathore}\ \emph {et~al.}(2005)\citenamefont
  {Rathore}, \citenamefont {Chopra},\ and\ \citenamefont
  {de~Pablo}}]{rathore05jcp}%
  \BibitemOpen
  \bibfield  {author} {\bibinfo {author} {\bibfnamefont {N.}~\bibnamefont
  {Rathore}}, \bibinfo {author} {\bibfnamefont {M.}~\bibnamefont {Chopra}}, \
  and\ \bibinfo {author} {\bibfnamefont {J.~J.}\ \bibnamefont {de~Pablo}},\
  }\href {\doibase 10.1063/1.1831273} {\bibfield  {journal} {\bibinfo
  {journal} {J. Chem. Phys.}\ }\textbf {\bibinfo {volume} {122}},\ \bibinfo
  {pages} {024111} (\bibinfo {year} {2005})}\BibitemShut {NoStop}%
\bibitem [{\citenamefont {Katzgraber}\ \emph {et~al.}(2006)\citenamefont
  {Katzgraber}, \citenamefont {Trebst}, \citenamefont {Huse},\ and\
  \citenamefont {Troyer}}]{katzgraber06jsm}%
  \BibitemOpen
  \bibfield  {author} {\bibinfo {author} {\bibfnamefont {H.~G.}\ \bibnamefont
  {Katzgraber}}, \bibinfo {author} {\bibfnamefont {S.}~\bibnamefont {Trebst}},
  \bibinfo {author} {\bibfnamefont {D.~A.}\ \bibnamefont {Huse}}, \ and\
  \bibinfo {author} {\bibfnamefont {M.}~\bibnamefont {Troyer}},\ }\href
  {\doibase 10.1088/1742-5468/2006/03/P03018} {\bibfield  {journal} {\bibinfo
  {journal} {J. Stat. Mech.: Theory Exp.}\ }\textbf {\bibinfo {volume}
  {2006}},\ \bibinfo {pages} {P03018} (\bibinfo {year} {2006})}\BibitemShut
  {NoStop}%
\bibitem [{\citenamefont {Patriksson}\ and\ \citenamefont {van~der
  Spoel}(2008)}]{patriksson08pccp}%
  \BibitemOpen
  \bibfield  {author} {\bibinfo {author} {\bibfnamefont {A.}~\bibnamefont
  {Patriksson}}\ and\ \bibinfo {author} {\bibfnamefont {D.}~\bibnamefont
  {van~der Spoel}},\ }\href {\doibase 10.1039/B716554D} {\bibfield  {journal}
  {\bibinfo  {journal} {Phys. Chem. Chem. Phys.}\ }\textbf {\bibinfo {volume}
  {10}},\ \bibinfo {pages} {2073} (\bibinfo {year} {2008})}\BibitemShut
  {NoStop}%
\bibitem [{\citenamefont {Guidetti}\ \emph {et~al.}(2012)\citenamefont
  {Guidetti}, \citenamefont {Rolando},\ and\ \citenamefont
  {Tripiccione}}]{guidetti2012jcp}%
  \BibitemOpen
  \bibfield  {author} {\bibinfo {author} {\bibfnamefont {M.}~\bibnamefont
  {Guidetti}}, \bibinfo {author} {\bibfnamefont {V.}~\bibnamefont {Rolando}}, \
  and\ \bibinfo {author} {\bibfnamefont {R.}~\bibnamefont {Tripiccione}},\
  }\href {\doibase 10.1016/j.jcp.2011.10.019} {\bibfield  {journal} {\bibinfo
  {journal} {J. Comput. Phys.}\ }\textbf {\bibinfo {volume} {231}},\ \bibinfo
  {pages} {1524} (\bibinfo {year} {2012})}\BibitemShut {NoStop}%
\bibitem [{\citenamefont {Ballard}\ and\ \citenamefont
  {Wales}(2014)}]{ballard14jctc}%
  \BibitemOpen
  \bibfield  {author} {\bibinfo {author} {\bibfnamefont {A.~J.}\ \bibnamefont
  {Ballard}}\ and\ \bibinfo {author} {\bibfnamefont {D.~J.}\ \bibnamefont
  {Wales}},\ }\href {\doibase 10.1021/ct500797a} {\bibfield  {journal}
  {\bibinfo  {journal} {J. Chem. Theory Comput. (JCTC)}\ }\textbf {\bibinfo
  {volume} {10}},\ \bibinfo {pages} {5599} (\bibinfo {year}
  {2014})}\BibitemShut {NoStop}%
\bibitem [{\citenamefont {Hsu}\ and\ \citenamefont
  {Grassberger}(2011)}]{hsugrassberger11jsp}%
  \BibitemOpen
  \bibfield  {author} {\bibinfo {author} {\bibfnamefont {H.-P.}\ \bibnamefont
  {Hsu}}\ and\ \bibinfo {author} {\bibfnamefont {P.}~\bibnamefont
  {Grassberger}},\ }\href {\doibase 10.1007/s10955-011-0268-x} {\bibfield
  {journal} {\bibinfo  {journal} {J. Stat. Phys.}\ }\textbf {\bibinfo {volume}
  {144}},\ \bibinfo {pages} {597} (\bibinfo {year} {2011})}\BibitemShut
  {NoStop}%
\bibitem [{\citenamefont {Hansmann}\ \emph {et~al.}(1996)\citenamefont
  {Hansmann}, \citenamefont {Okamoto},\ and\ \citenamefont
  {Eisenmenger}}]{hansmann96cpl}%
  \BibitemOpen
  \bibfield  {author} {\bibinfo {author} {\bibfnamefont {U.~H.~E.}\
  \bibnamefont {Hansmann}}, \bibinfo {author} {\bibfnamefont {Y.}~\bibnamefont
  {Okamoto}}, \ and\ \bibinfo {author} {\bibfnamefont {F.}~\bibnamefont
  {Eisenmenger}},\ }\href {\doibase 10.1016/0009-2614(96)00761-0} {\bibfield
  {journal} {\bibinfo  {journal} {Chem. Phys. Lett.}\ }\textbf {\bibinfo
  {volume} {259}},\ \bibinfo {pages} {321} (\bibinfo {year}
  {1996})}\BibitemShut {NoStop}%
\bibitem [{\citenamefont {Junghans}\ \emph {et~al.}(2014)\citenamefont
  {Junghans}, \citenamefont {Perez},\ and\ \citenamefont
  {Vogel}}]{junghans14jctc}%
  \BibitemOpen
  \bibfield  {author} {\bibinfo {author} {\bibfnamefont {C.}~\bibnamefont
  {Junghans}}, \bibinfo {author} {\bibfnamefont {D.}~\bibnamefont {Perez}}, \
  and\ \bibinfo {author} {\bibfnamefont {T.}~\bibnamefont {Vogel}},\ }\href
  {\doibase 10.1021/ct500077d} {\bibfield  {journal} {\bibinfo  {journal} {J.
  Chem. Theory Comput. (JCTC)}\ }\textbf {\bibinfo {volume} {10}},\ \bibinfo
  {pages} {1843} (\bibinfo {year} {2014})}\BibitemShut {NoStop}%
\bibitem [{\citenamefont {Laio}\ and\ \citenamefont
  {Parrinello}(2002)}]{laioparrinello02pnas}%
  \BibitemOpen
  \bibfield  {author} {\bibinfo {author} {\bibfnamefont {A.}~\bibnamefont
  {Laio}}\ and\ \bibinfo {author} {\bibfnamefont {M.}~\bibnamefont
  {Parrinello}},\ }\href {\doibase 10.1073/pnas.202427399} {\bibfield
  {journal} {\bibinfo  {journal} {Proc. Natl. Acad. Sci.}\ }\textbf {\bibinfo
  {volume} {99}},\ \bibinfo {pages} {12562} (\bibinfo {year}
  {2002})}\BibitemShut {NoStop}%
\bibitem [{\citenamefont {Dama}\ \emph {et~al.}(2014)\citenamefont {Dama},
  \citenamefont {Parrinello},\ and\ \citenamefont {Voth}}]{dama14prl}%
  \BibitemOpen
  \bibfield  {author} {\bibinfo {author} {\bibfnamefont {J.~F.}\ \bibnamefont
  {Dama}}, \bibinfo {author} {\bibfnamefont {M.}~\bibnamefont {Parrinello}}, \
  and\ \bibinfo {author} {\bibfnamefont {G.~A.}\ \bibnamefont {Voth}},\ }\href
  {\doibase 10.1103/PhysRevLett.112.240602} {\bibfield  {journal} {\bibinfo
  {journal} {Phys. Rev. Lett.}\ }\textbf {\bibinfo {volume} {112}},\ \bibinfo
  {pages} {240602} (\bibinfo {year} {2014})}\BibitemShut {NoStop}%
\bibitem [{\citenamefont {Kim}\ \emph {et~al.}(2006)\citenamefont {Kim},
  \citenamefont {Straub},\ and\ \citenamefont {Keyes}}]{kim06prl}%
  \BibitemOpen
  \bibfield  {author} {\bibinfo {author} {\bibfnamefont {J.}~\bibnamefont
  {Kim}}, \bibinfo {author} {\bibfnamefont {J.~E.}\ \bibnamefont {Straub}}, \
  and\ \bibinfo {author} {\bibfnamefont {T.}~\bibnamefont {Keyes}},\ }\href
  {\doibase 10.1103/PhysRevLett.97.050601} {\bibfield  {journal} {\bibinfo
  {journal} {Phys. Rev. Lett.}\ }\textbf {\bibinfo {volume} {97}},\ \bibinfo
  {pages} {050601} (\bibinfo {year} {2006})}\BibitemShut {NoStop}%
\bibitem [{\citenamefont {Geyer}(1991)}]{geyer91proc}%
  \BibitemOpen
  \bibfield  {author} {\bibinfo {author} {\bibfnamefont {C.~J.}\ \bibnamefont
  {Geyer}},\ }in\ \href@noop {} {\emph {\bibinfo {booktitle} {Computing Science
  and Statistics: Proceedings of the 23rd Symposium on the Interface}}},\
  \bibinfo {editor} {edited by\ \bibinfo {editor} {\bibfnamefont {E.~M.}\
  \bibnamefont {Keramidas}}}\ (\bibinfo  {publisher} {Interface Foundation},\
  \bibinfo {address} {Fairfax Station, VA},\ \bibinfo {year} {1991})\ p.\
  \bibinfo {pages} {156}\BibitemShut {NoStop}%
\bibitem [{\citenamefont {Lyubartsev}\ \emph {et~al.}(1992)\citenamefont
  {Lyubartsev}, \citenamefont {Martsinovski}, \citenamefont {Shevkunov},\ and\
  \citenamefont {Vorontsov-Velyaminov}}]{lyubartsev92jcp}%
  \BibitemOpen
  \bibfield  {author} {\bibinfo {author} {\bibfnamefont {A.~P.}\ \bibnamefont
  {Lyubartsev}}, \bibinfo {author} {\bibfnamefont {A.~A.}\ \bibnamefont
  {Martsinovski}}, \bibinfo {author} {\bibfnamefont {S.~V.}\ \bibnamefont
  {Shevkunov}}, \ and\ \bibinfo {author} {\bibfnamefont {P.~N.}\ \bibnamefont
  {Vorontsov-Velyaminov}},\ }\href {\doibase 10.1063/1.462133} {\bibfield
  {journal} {\bibinfo  {journal} {J. Chem. Phys.}\ }\textbf {\bibinfo {volume}
  {96}},\ \bibinfo {pages} {1776} (\bibinfo {year} {1992})}\BibitemShut
  {NoStop}%
\bibitem [{\citenamefont {Wang}\ and\ \citenamefont
  {Landau}(2001)}]{wanglandau01prl}%
  \BibitemOpen
  \bibfield  {author} {\bibinfo {author} {\bibfnamefont {F.}~\bibnamefont
  {Wang}}\ and\ \bibinfo {author} {\bibfnamefont {D.~P.}\ \bibnamefont
  {Landau}},\ }\href {\doibase 10.1103/PhysRevLett.86.2050} {\bibfield
  {journal} {\bibinfo  {journal} {Phys. Rev. Lett.}\ }\textbf {\bibinfo
  {volume} {86}},\ \bibinfo {pages} {2050} (\bibinfo {year}
  {2001})}\BibitemShut {NoStop}%
\bibitem [{\citenamefont {Vogel}\ and\ \citenamefont
  {Perez}(2015)}]{vogelperez2015prl}%
  \BibitemOpen
  \bibfield  {author} {\bibinfo {author} {\bibfnamefont {T.}~\bibnamefont
  {Vogel}}\ and\ \bibinfo {author} {\bibfnamefont {D.}~\bibnamefont {Perez}},\
  }\href {\doibase 10.1103/PhysRevLett.115.190602} {\bibfield  {journal}
  {\bibinfo  {journal} {Phys. Rev. Lett.}\ }\textbf {\bibinfo {volume} {115}},\
  \bibinfo {pages} {190602} (\bibinfo {year} {2015})}\BibitemShut {NoStop}%
\bibitem [{\citenamefont {Darve}\ \emph {et~al.}(2008)\citenamefont {Darve},
  \citenamefont {Rodr{\'i}guez-G{\'o}mez},\ and\ \citenamefont
  {Pohorille}}]{darve08jcp}%
  \BibitemOpen
  \bibfield  {author} {\bibinfo {author} {\bibfnamefont {E.}~\bibnamefont
  {Darve}}, \bibinfo {author} {\bibfnamefont {D.}~\bibnamefont
  {Rodr{\'i}guez-G{\'o}mez}}, \ and\ \bibinfo {author} {\bibfnamefont
  {A.}~\bibnamefont {Pohorille}},\ }\href {\doibase 10.1063/1.2829861}
  {\bibfield  {journal} {\bibinfo  {journal} {J. Chem. Phys.}\ }\textbf
  {\bibinfo {volume} {128}},\ \bibinfo {pages} {144120} (\bibinfo {year}
  {2008})}\BibitemShut {NoStop}%
\bibitem [{Note1()}]{Note1}%
  \BibitemOpen
  \bibinfo {note} {The management of all global data is done by an additional
  master process which does not perform a MD run himself.}\BibitemShut {Stop}%
\bibitem [{\citenamefont {Nadler}\ and\ \citenamefont
  {Hansmann}(2007)}]{nadler07pre}%
  \BibitemOpen
  \bibfield  {author} {\bibinfo {author} {\bibfnamefont {W.}~\bibnamefont
  {Nadler}}\ and\ \bibinfo {author} {\bibfnamefont {U.~H.~E.}\ \bibnamefont
  {Hansmann}},\ }\href {\doibase 10.1103/PhysRevE.75.026109} {\bibfield
  {journal} {\bibinfo  {journal} {Phys. Rev. E}\ }\textbf {\bibinfo {volume}
  {75}},\ \bibinfo {pages} {026109} (\bibinfo {year} {2007})}\BibitemShut
  {NoStop}%
\bibitem [{\citenamefont {Bittner}\ \emph {et~al.}(2008)\citenamefont
  {Bittner}, \citenamefont {Nu{\ss}baumer},\ and\ \citenamefont
  {Janke}}]{elmar08prl}%
  \BibitemOpen
  \bibfield  {author} {\bibinfo {author} {\bibfnamefont {E.}~\bibnamefont
  {Bittner}}, \bibinfo {author} {\bibfnamefont {A.}~\bibnamefont
  {Nu{\ss}baumer}}, \ and\ \bibinfo {author} {\bibfnamefont {W.}~\bibnamefont
  {Janke}},\ }\href {\doibase 10.1103/PhysRevLett.101.130603} {\bibfield
  {journal} {\bibinfo  {journal} {Phys. Rev. Lett.}\ }\textbf {\bibinfo
  {volume} {101}},\ \bibinfo {pages} {130603} (\bibinfo {year}
  {2008})}\BibitemShut {NoStop}%
\bibitem [{\citenamefont {Williams}\ \emph {et~al.}(2006)\citenamefont
  {Williams}, \citenamefont {Mishin},\ and\ \citenamefont
  {Hamilton}}]{williams06msmse}%
  \BibitemOpen
  \bibfield  {author} {\bibinfo {author} {\bibfnamefont {P.~L.}\ \bibnamefont
  {Williams}}, \bibinfo {author} {\bibfnamefont {Y.}~\bibnamefont {Mishin}}, \
  and\ \bibinfo {author} {\bibfnamefont {J.~C.}\ \bibnamefont {Hamilton}},\
  }\href {\doibase 10.1088/0965-0393/14/5/002} {\bibfield  {journal} {\bibinfo
  {journal} {Model. Simul. Mater. Sci. Eng.}\ }\textbf {\bibinfo {volume}
  {14}},\ \bibinfo {pages} {817} (\bibinfo {year} {2006})}\BibitemShut
  {NoStop}%
\bibitem [{Note2()}]{Note2}%
  \BibitemOpen
  \bibinfo {note} {If there are no measured values in particular energy bins,
  we simply copy the closest valid value from a lower energy.}\BibitemShut
  {Stop}%
\bibitem [{Note3()}]{Note3}%
  \BibitemOpen
  \bibinfo {note} {This behavior is known and simply illustrates the perils of
  choosing a reaction variable that involves first-order-like
  transitions.}\BibitemShut {Stop}%
\end{thebibliography}%

\end{document}